\documentclass[sigconf,natbib=true,anonymous=False]{acmart}
\usepackage{tcolorbox}
\usepackage{listings}
\tcbuselibrary{listingsutf8} 
\usepackage{url}
\usepackage{multicol}
\usepackage{multirow}
\usepackage{float}        
\usepackage{tabularray}
\AtBeginDocument{%
  }

\setcopyright{acmlicensed}
\copyrightyear{2018}
\acmYear{2018}
\begin{document}

\title{OrdRankBen: A Novel Ranking Benchmark for Ordinal Relevance in NLP}

\author{Yan Wang}
\authornotemark[1]
\affiliation{%
  \institution{The Fin AI}
  \country{USA}
}
\email{wy2266336@gmail.com}

\author{Lingfei Qian}
\authornote{Both authors contributed equally to this research.}
\affiliation{%
  \institution{The Fin AI}
  \country{USA}
}
\email{lfqian94@gmail.com}

\author{Xueqing Peng}
\affiliation{%
  \institution{The Fin Ai}
  \country{USA}}
\email{xueqing.peng2023@gmail.com}

\author{Jimin Huang}
\authornotemark[2]
\affiliation{%
  \institution{The Fin Ai}
  \country{USA}}
\email{jimin.huang@thefin.ai}

\author{Dongji Feng}
\authornote{Corresponding Authors}
\affiliation{%
    \institution{Gustavus Adolphus College}
    \city{St.Peter}
    \state{Minnesota}
    \country{USA}
}
\email{djfeng@gustavus.edu}






\renewcommand{\shortauthors}{Yan et al.}

\begin{abstract} 
The evaluation of ranking tasks remains a significant challenge in natural language processing (NLP), particularly due to the lack of direct labels for results in real-world scenarios. Benchmark datasets play a crucial role in providing standardized testbeds that ensure fair comparisons, enhance reproducibility, and enable progress tracking, facilitating rigorous assessment and continuous improvement of ranking models.
Existing NLP ranking benchmarks typically use binary relevance labels or continuous relevance scores, neglecting ordinal relevance scores. However, binary labels oversimplify relevance distinctions, while continuous scores lack a clear ordinal structure, making it challenging to capture nuanced ranking differences effectively.
To address these challenges, we introduce \textbf{OrdRankBen}, a novel benchmark designed to capture multi-granularity relevance distinctions. Unlike conventional benchmarks, OrdRankBen incorporates structured ordinal labels, enabling more precise ranking evaluations.
Given the absence of suitable datasets for ordinal relevance ranking in NLP, we constructed two datasets with distinct ordinal label distributions. We further evaluate various models for three model types: ranking-based language models, general large language models, and ranking-focused large language models on these datasets
Experimental results show that ordinal relevance modeling provides a more precise evaluation of ranking models, improving their ability to distinguish multi-granularity differences among ranked items—crucial for tasks that demand fine-grained relevance differentiation. 
\end{abstract}




\maketitle

\section{Introduction}

Ranking methods have been extensively studied as machine learning problems due to their significance in various tasks~\cite{huang2020embedding,lin2021pyserini,lin2024mixed,gao2021unsupervised,wang2022text}. With the introduction of the learning-to-rank (LTR) approaches~\cite {cao2007learning,liu2009learning,qin2010letor}, ranking models leverage supervised machine learning techniques for optimization, broadly categorized into pointwise, pairwise, and listwise methods~\cite{cao2007learning}. These techniques improve ranking quality by directly optimizing evaluation metrics such as Mean Average Precision (MAP)\cite{zuva2012evaluation}, Normalized Discounted Cumulative Gain (NDCG)\cite{webber2010similarity}, and Expected Reciprocal Rank (ERR)~\cite{chapelle2009expected}.

To evaluate ranking models effectively, machine learning ranking datasets employ different types of relevance scores, such as binary (e.g., relevant or not relevant), continuous (e.g., floating-point scores representing degrees of relevance), and ordinal (e.g., structured labels capturing ordered relevance levels). These diverse scoring schemes enable comprehensive benchmarking, testing a model's ability to capture different relevance distinctions across different ranking scenarios.



In natural language processing (NLP) domain, ranking tasks play a crucial role in research and have led to the development of extensive benchmark resources such as MSMARCO\cite{bajaj2016ms}, KILT\cite{petroni2020kilt}, BEIR\cite{thakur2021beir}, FollowIR\cite{weller2024followir}, InstructIR\cite{oh2024instructir}, and MAIR\cite{sun2024mair}. These benchmarks are essential for providing standardized datasets and evaluation metrics, ensuring fair comparison and reproducibility across different approaches.

However, these benchmark datasets primarily use binary labels or continuous relevance scores for ranking metrics, overlooking the significance of ordinal relevance scores. This limitation hinders their ability to capture multi-granularity relevance differentiation and poses challenges in real-world ranking scenarios. For example, \textbf{binary labels} oversimplify the ranking problem by failing to capture fine-grained differences between items, which is essential for search engines that must distinguish results with varying levels of informativeness and relevance~\cite{zhuang2024beyond}. \textbf{Continuous relevance scores}, while offering more subtle differences, often lack an explicit ordinal structure, making it difficult to model the relative importance of items in a ranked list, which is an essential requirement for personalized recommendations~\cite{cao2007learning}. Additionally, many standard ranking metrics in NLP do not explicitly account for ordinal relationships, which can degrade performance in tasks that require a more precise relevance ranking~\cite{karpukhin2020dense,zamani2021neural}.

To bridge this gap, we introduce OrdRankBen, a new benchmark for ordinal relevance ranking in the NLP domain. Unlike traditional binary or continuous scoring methods, \textbf{OrdRankBen incorporates structured ordinal labels and features two datasets with distinct label distributions}, allowing for a more comprehensive evaluation of ranking models. This setup helps models better capture multi-granularity differences in relevance. Additionally, we include evaluation metrics for ordinal ranking, providing a more robust assessment of ranking-based language models (LMs), general large language models (LLMs), and ranking-focused large language models (LLMs) in tasks where precise relevance differentiation is essential. We have released the OrdRankBen dataset and evaluation methods on GitHub\footnote{\url{https://github.com/Yan2266336/OrdRankBen}}.

The major contributions of this work include:
\begin{itemize}
    \item We introduce OrdRankBen, a new ranking benchmark for the NLP domain, based on ordinal relevance scores. This comprehensive benchmark includes two datasets with distinct ordinal label distributions, enabling a multifaceted evaluation of LLMs' ranking capabilities.

    \item To comprehensively evaluate the ranking capabilities of existing NLP models, we examined nine models across three categories: Ranking-based LMs, general purpose LLMs, and ranking-specific LLMs. Our evaluation analyzes their performance on the ranking task with ordinal relevance labels, providing insights into their effectiveness in handling fine-grained relevance distinctions.

    \item For better integration with ranking in the natural language domain and a more comprehensive evaluation of models in this field, we released two versions of the OrdRankBen ranking dataset: one tailored for LM-series models and the other for LLM-series models. 
\end{itemize}














\section{OrdRankBen}
In this section, we present OrdRankBen, an NLP ranking benchmark that extends traditional binary and continuous relevance scoring to an ordinal relevance framework. We first outline the task formulation (Section~\ref{sec:formulation}), then detail the data collection process (Section~\ref{sec:collection}), and finally, describe the annotation process (Section~\ref{sec:annotation}).

\subsection{Task Formulation}
\label{sec:formulation}
We formally define the task of OrdRankBen in the context of language models as follows: Given a query $q$ and a text set $T=\{t_1, t_2, \cdots, t_k\}$ of $k$ candidate texts related to $q$, each candidate text $t_i$ is assigned an ordinal relevance label $s$, where $s \in \{1,2,\cdots,m\}$ and $m$ denotes the maximum value on the ordinal scale. Our goal is to leverage various LMs to rank these candidate texts based on the semantic similarity score $score_i$ between $q$ and $t_i$, and evaluate the ranking capability of these LMs by analyzing the relationship between the priority of the ranked texts and their corresponding ordinal relevance labels $s$.

\begin{equation}
    score_i = f_{LM}(q,t_i)
\end{equation}

\begin{equation}
    R = \mathrm{Rank}(T,score_i)
\end{equation}
where $f_{LM}(\cdot)$ is a language model that computes a semantic similarity score $score_i$ for each candidate text $t_i$, $R=\{t_{\pi_1},t_{\pi_2},\cdots,t_{\pi_k}\}$ is the ranked list of candidate text ordered by $score_i$.

\subsection{Data collection}
\label{sec:collection}

To construct the evaluation dataset for the OrdRankBen benchmark, we sourced data from the MicroSoft MAchine Reading COmprehension (MSMARCO)\footnote{\url{https://microsoft.github.io/msmarco/Datasets}}, a large-scale dataset designed for machine reading comprehension. MSMARCO encompasses various NLP and IR tasks, including question answering, passage ranking, document ranking, keyphrase extraction, and conversational search.

For our benchmark, we focus on two ranking tasks: document ranking and passage ranking. The statistical information is presented in Table~\ref{tab:statistic}.

\begin{table}\small
\small
    \caption{Basic statistical information about document and passage ranking tasks in the MSMARCO dataset.}
    \centering
    \begin{tabular}{llccc}
    \hline
         Tasks & Items & Training & Validation & Testing \\ \hline
        \multirow{3}*{Document} & \# Query & 367,013 & 5,193 & 5,793 \\
        ~ & \# Text & 36,701,116 & 519,300 & 579,300 \\
        ~ & Relevance type & \multicolumn{3}{c}{continuous relevance} \\ \hline
        \multirow{3}*{Passage} & \# Query & 532,761 & 59,273 & - \\
        ~ & \# Text & 478,002,393 & 6,668,967 & - \\
        ~ & Relevance type & \multicolumn{3}{c}{binary relevance} \\ \hline
    \end{tabular}
    \label{tab:statistic}
\end{table}

The document ranking task involves re-ranking an initial set of 100 documents retrieved by a basic IR system based on their relevance to the query question. Similarly, the passage ranking task aims to reorder the initial set of 1,000 passages based on how likely they are to contain the correct answer. The document ranking dataset has a training, validation, and testing set, but the passage ranking dataset does not include a testing set. 

\subsection{Data Annotation}
\label{sec:annotation}

To better indicate the ordinal relevance labels in our benchmark, we define the ordinal scale for relevance labels $s$ as $\{1,2,3,4,5\}$, which represents the different degrees of relevance between query question $q$ and candidate text $t_i$. The detailed definitions of these labels are provided in Table~\ref{tab:definition}. A score of 1 indicates no relevance (0\%), 2 represents slight relevance (25\%), 3 denotes moderate relevance (50\%), 4 signifies high relevance (75\%), and 5 indicates full relevance (100\%) between the query and the candidate text.

\begin{table}\small
\small
  \caption{The definitions and statistical information of relevance label on OrdRankBen benchmark. Note: `DR' indicates the document ranking task, and `PR' indicates the passage ranking task.}
  \label{tab:definition}
  \begin{tabular}{lccccc}
    \toprule
    Relevance & 0\%  & 25\%  & 50\%  & 75\%  & 100\% \\
    \midrule
    Label & 1 & 2 & 3 &4 & 5 \\
    \midrule
    \# Label in DR & 115,860 & 115,8601 & 115,860 & 115,860 & 115,860 \\
    \# Label in PR & 22,021 & 10,893 & 84,170 & 161,019 & 399,997 \\
  \bottomrule
\end{tabular}
\end{table}

Specifically, we re-annotated 678,100 query-passage pairs from the MSMARCO validation set for the passage ranking task and 579,300 query-document pairs from the MSMARCO testing set for the document ranking task, assigning each pair a relevance label $s$. The statistical distribution of these labels is presented in Table~\ref{tab:definition}. Due to the differences in the original relevance types of these two tasks, we adopt the following two approaches to assign ordinal relevance labels separately for each task.

\textbf{Uniform selection for document ranking task}: For MSMARCO testing set in the document ranking task, we apply a uniform selection strategy to re-annotate each query-document pair with an ordinal relevance label for our document ranking task. In the original testing set, each query has 100 candidate documents, and these candidate documents are ranked by their continuous (log-likelihood) relevance to the query. To create ordinal labels for our task, we divide these candidates into five groups of 20 in descending order: the top 20 are labeled 5, the next 20 are labeled 4, and so on, with the last 20 labeled 1. As shown in Table~\ref{tab:definition}, this method ensures an equal label distribution, resulting in a balanced dataset for the document ranking task.

\textbf{GPT judgment for passage ranking task}: Following the MSMARCO passage ranking task, we constructed our evaluation dataset from its validation set. To ensure consistency in the number of candidates with the document ranking task—where each query has 100 candidate documents—we randomly selected 6,781 queries with more than 100 candidate passages. Based on the provided binary relevance scores, passages with a score of 1 were assigned an ordinal relevance label of 5 for each selected query. For the remaining candidates, we randomly selected a subset of passages with a binary relevance score of 0 to ensure a total of 100 candidates per query, consisting of both relevant and non-relevant passages. Finally, we utilized the GPT model to assign ordinal relevance labels to the non-relevant passages. Notably, to maintain fairness, we allowed the GPT model to assign an ordinal relevance label of 5 to some non-relevant passages if deemed appropriate. The prompt template of this method is shown in Appendix~\ref{template}.





\section{Experiment Setup}

\subsection{Evaluation metrics}
To better evaluate LLMs' ranking capability, we chose ERR and nDCG as evaluation metrics. \textbf{These metrics effectively handle ordinal relevance labels and provide a comprehensive assessment of ranking performance}. The formulation of ERR and nDCG is shown in Appendix~\ref{metrics}.

\textbf{ERR} incorporates a probabilistic relevance distribution, where higher ordinal labels contribute more significantly to the final score, ensuring that the ranking evaluation reflects the varying degrees of relevance.

\textbf{nDCG} applies graded relevance weighting with a logarithmic discount, ensuring higher relevance labels have greater influence while accounting for positional effects. By directly integrating ordinal relevance scores without thresholding or binarization, it effectively preserves relevance distinctions, making it a robust metric for assessing global ranking quality, especially in longer ranked lists.

\subsection{Evaluated language models} 
To comprehensively evaluate the ranking capabilities of various language models (LMs), we selected 9 models spanning three categories: (1) traditional ranking-based LMs: CoRom~\cite{zhang2022hlatrenhancemultistagetext}, RankingBERT~\cite{nogueira2020passagererankingbert}, and RankT5~\cite{zhuang2023rankt5}; (2) state-of-the-art (SOTA) general LLMs: GPT-4o-mini~\cite{hurst2024gpt}, 
Gemma-2-27-it~\cite{team2024gemma}, and LLaMA-3.3-70b-Instruct~\cite{dubey2024llama}; and (3) ranking-focused LLMs: Gte-Qwen2-7B-instruct~\cite{li2023towards}, Stella-1.5B~\cite{zhang2025jasperstelladistillationsota}, and RankLLaMA~\cite{rankllama}. These models were tested on passage and document ranking tasks from the OrdRankBen benchmark. Table~\ref{tab:models} details the specific versions used. For consistency, we introduce abbreviations for each model throughout the paper.

\begin{table}\small
  \caption{Details of evaluated models. Note: $\sim$ denotes approximately. All ranking-based and ranking-focused models are fine-tuned on the raw MSMARCO dataset.}
  \label{tab:models}
  \begin{tabular}{lll}
    \toprule
    Category & Model name & Short form  \\
    \midrule
    \multirow{3}*{Ranking-based LMs}& CoRom-base & CoRom  \\
    ~ & RankingBERT & RBert\\
    ~ & RankT5-base &RT5\\
    \midrule
    \multirow{4}*{General LLMs}& GPT-4o-mini & GPTmini\\
    ~ & Gemma-2-27b-it &Gemma \\
    ~ & LLaMA-3.3-70b-Instruct & LInstruct \\
    \midrule
    \multirow{3}*{Ranking-focused LLMs}& gte-Qwen2-7B-instruct &GteQwen2\\
    ~ & stella\_en\_1.5B\_v5  & Stella \\
    ~ & rankllama & RLLaMA \\
  \bottomrule
\end{tabular}
\end{table}

\section{Experimental Results}

\begin{table*}[ht]\small
  \caption{Relative ERR performance of different Ranking Methods for cutoff \textit{k} on two tasks}
  \label{tab:err_performance2}
  \centering
  \begin{tabular}{lcccccc||cccccc}
    \hline
    \multirow{3}*{Model Name} & \multicolumn{6}{c||}{ Document ranking (DR)} & \multicolumn{6}{c}{Passage ranking (PR)} \\ 
    \cline{2-13}
    &\multicolumn{6}{c||}{ERR@} & \multicolumn{6}{c}{ERR@}\\
    
    \cline{2-13}
    ~ & 5  & 10 & 15 & 20 & 30  & 60 & 5  & 10 & 15& 20 & 30 & 60 \\
    
    \hline
    CoRom & 0.90966 (5)& 0.91020& 0.91022& 0.91022& 0.91022& 0.91022& 0.91348 (5)& 0.91549& 0.91589& 0.91603& 0.91612& 0.91615\\
    RBert & 0.92677 (4)& 0.92710& 0.92711& 0.92711& 0.92711& 0.92711& 0.89737 (7)& 0.89952& 0.89996& 0.90010& 0.90018& 0.90021\\
    RT5 & 0.90639 (6)& 0.90684& 0.90684& 0.90684& 0.90684& 0.90684& 0.86948 (8)& 0.87173& 0.87214& 0.87228& 0.87236& 0.87239\\
    \hline
    GPTmini & 0.98035 (1)& 0.98049& 0.98050& 0.98050& 0.98050& 0.98050& 0.91756 (2)& 0.91929& 0.91962& 0.91973& 0.91979& 0.91982\\
    Gemma & 0.89131 (7) & 0.89200& 0.89201& 0.89201& 0.89201& 0.89201& 0.74057 (9)& 0.74866& 0.74987& 0.75015& 0.75030& 0.75035\\
    LInstruct & 0.94968 (2)& 0.94998 & 0.94998& 0.94998& 0.94998& 0.94998& 0.90862 (6)& 0.91068& 0.91102& 0.91114& 0.91120& 0.91123\\
    \hline
    GteQwen2 & 0.82285 (8)& 0.82387& 0.82388& 0.82388& 0.82388& 0.82388& 0.91569 (4)& 0.91743& 0.91780& 0.91793& 0.91800& 0.91803\\
    Stella & 0.74454 (9) & 0.74759& 0.74770& 0.74771& 0.74771& 0.74772& 0.91585 (3)& 0.91755& 0.91790& 0.91802& 0.91809& 0.91812\\
    RLLaMA & 0.93542 (3)& 0.93585& 0.93586& 0.93586& 0.93586& 0.93586& 0.92318 (1)& 0.92477& 0.92511& 0.92522& 0.92530& 0.92532\\
    \hline
  \end{tabular}
\end{table*}

\begin{table*}\small
  \caption{Relative nDCG performance of different Ranking Methods for cutoff \textit{k} on two tasks}
  \label{tab:nDCG_performance2}
  \centering
  \begin{tabular}{lcccccc||cccccc}
    \hline
    \multirow{3}*{Model Name} & \multicolumn{6}{c||}{Document ranking (DR)} & \multicolumn{6}{c}{Passage ranking (PR)} \\ \cline{2-13}

    &\multicolumn{6}{c||}{nDCG@} & \multicolumn{6}{c}{ nDCG@}\\

    \cline{2-13}
    
    ~ & 5 & 10 & 15& 20& 30 & 60 & 5 & 10 & 15& 20 & 30 & 60\\
    \hline
    CoRom & 0.65604 (6)& 0.59576 (7)& 0.55684 (7) & 0.52986 (7) & 0.56089 (7) & 0.65686 (7)& 0.79440 (6)& 0.78898& 0.79107& 0.79651& 0.81089& 0.85760\\
    RBert & 0.71336 (4) & 0.66224 (4)& 0.62470 (5)& 0.59549(5)& 0.62732(4) & 0.70710 (4)& 0.77278 (7)& 0.77173& 0.77728& 0.78426& 0.80014& 0.84729\\
    RT5 & 0.68077 (5)& 0.63891 (5)& 0.60597(6)& 0.57931(6)& 0.61209 (6)& 0.69500 (6)& 0.75125 (8)& 0.75854& 0.76508& 0.77299& 0.78965& 0.83788\\
    \hline
    GPTmini & 0.85119 (1)& 0.78061 (1)& 0.72987 (1)& 0.68985 (1)& 0.71985 (2)& 0.79571 (2)& 0.82903 (1)& 0.83203& 0.83669& 0.84275& 0.85540& 0.89042\\
    Gemma & 0.61073 (7)& 0.60421 (6)& 0.63132 (3)& 0.66708 (2)& 0.78573  (1)& 0.87361 (1)& 0.52895 (9)& 0.55290& 0.57041& 0.58414& 0.60870& 0.67936\\
    LInstruct& 0.75286 (2) & 0.69313 (2)& 0.65174 (2)& 0.62103 (3)& 0.65690 (3)& 0.75610 (3)& 0.80788 (3)& 0.81149& 0.81732& 0.82392& 0.83779& 0.87608\\
    \hline
    GteQwen2 & 0.51261 (8)& 0.48809 (8)& 0.47240(8)& 0.46072(8)& 0.50342(8)& 0.61576(8)& 0.80338 (5)& 0.79881& 0.80161& 0.80693& 0.82079& 0.86451\\
    Stella & 0.41474 (9)& 0.41650 (9)& 0.41489 (9)& 0.41312 (9)& 0.46550  (9)& 0.59318  (9)& 0.80487 (4)& 0.80169& 0.80424& 0.80995& 0.82408& 0.86711\\
    RLLaMA & 0.72930 (3) & 0.67127 (3)& 0.63023  (4)& 0.59852 (4)& 0.62662 (5)& 0.70517 (5)& 0.82371 (2)& 0.82046& 0.82295& 0.82770& 0.83980& 0.87886\\
    \hline
  \end{tabular}
\end{table*}

Table ~\ref{tab:err_performance2} and ~\ref{tab:nDCG_performance2} summarizes the average performances of different ranking methods induced by \textit{ERR@k} and \textit{nDCG@k} for various values of cutoff \textit{k}, i.e., k = [5, 10, 15, 20, 30, 60] for DR and PR tasks respectively. 
We first focus on the performance induced by \textit{ERR@k}. In Table ~\ref{tab:err_performance2}, we observe that GPTmini, a general LLM achieves the best for DR task, while RLLaMA, a Ranking-based LLM attains the best for PR tasks with \textit{ERR@k}, for all k various from 5 to 60. 
Meanwhile, the variance in the performance induced by \textit{ERR} with different cutoff k  by different ranking methods is small in both tasks, the average standard deviation of all ranking methods across all cutoff k is 0.0003 on DR task and 0.001 on PR task (Table ~\ref{tab:std_avg}).

For both tasks, the average \textit{ERR@k} for each ranking method maintains a non-decreasing trend as we raise k. For instance, in the DR task, Stella achieves an \textit{ERR} score of 0.74454 for k = 5 and 0.74772 for k = 60. One possible reason for this trend is that all ranking methods are capable of placing highly relevant candidates at the top, while items with zero / low relevance are either rare or absent. Another possible reason is that \textit{ERR} is less sensitive to low-relevance items, which means it may not accurately capture the variance in the data ~\cite{sakai2007properties}, resulting in the smallest standard deviation on both tasks.

We now focus on the average performance of different ranking methods induced by \textit{nDCG@k}. From Table ~\ref{tab:nDCG_performance2}, GPTmini obtains the best performance on both tasks, according to the average \textit{nDCG} score. Interestingly, unlike \textit{ERR@k},  \textit{nDCG@k} does not hold an increasing trend while we increase the cutoff value on both tasks. For instance, on the DR task,  RT5 achieves an \textit{nDCG} value of 0.68077 when k = 5, while an \textit{nDCG} value of 0.57931 when k = 20. One possible explanation is that \textit{nDCG} is sensitive to the cutoff, making it more responsive than \textit{ERR} in detecting differences in top-ranking results ~\cite{webber2010similarity,karmaker2020empirical}. Table ~\ref{tab:std_avg} presents the standard deviation of \textit{nDCG} scores for both tasks, which are 0.061 for DR and 0.028 for PR. These values are relatively higher compared to the standard deviations of \textit{ERR} for the same tasks.
Additionally, we report the average \textit{nDCG} and \textit{ERR} scores across all methods and cutoffs in the same Table. It is evident that \textit{nDCG} tends to penalize the performance of all ranking methods more severely than \textit{ERR}, resulting in comparatively lower scores. We compare the relative ranking of methods in terms of their $ERR$ and $nDCG$ scores, separately, \textit{within} a task or  \textit{across} different tasks (DR Vs PR). Details in Appendix ~\ref{appen:Relative Ranking} due to space.

\begin{table}[!htb]
\centering
\caption{Summary of the aggregate Standard Deviation (Std) and Average (Avg) performance derived from ERR/nDCG across all cutoff values for all ranking methods on two tasks.}
\label{tab:std_avg}

\begin{tabular}{ll|l|l}
\hline
\multicolumn{2}{l|}{}                           & Std    & Avg    \\ \hline
\multicolumn{1}{c|}{\multirow{2}{*}{ERR}}  & DR & 0.0003 & 0.8969 \\ \cline{2-4} 
\multicolumn{1}{c|}{}                      & PR & 0.001  & 0.8916 \\ \hline
\multicolumn{1}{l|}{\multirow{2}{*}{nDCG}} & DR & 0.061  & 0.6278 \\ \cline{2-4} 
\multicolumn{1}{l|}{}                      & PR & 0.028  & 0.7902 \\ \hline
\end{tabular}%
\end{table}

\section{Conclusion}
In this work, we present OrdRankBen, a novel NLP ranking benchmark that bridges the gap between binary and continuous relevance scoring by incorporating structured ordinal labels for more precise relevance differentiation and comprehensive model evaluation. 
We introduce two ranking task datasets—passage ranking and document ranking—with distinct ordinal relevance label distributions. Additionally, we evaluate nine language models across three categories to assess their ability to capture subtle ordinal relevance differences. Experimental results show that ranking based language models (LMs), trained on binary or continuous label datasets, exhibit average performance on our proposed tasks. In contrast, general LLMs, particularly GPTmini and LInstruct, demonstrate superior performance across both tasks. Our analysis indicates that 
the \textit{nDCG} metric is sensitive to the cutoff in our tasks and tends to impose a heavier penalty on the model performance compared to \textit{ERR}. 
Additionally, the incorporation of ordinal relevance labels in our evaluation framework plays a crucial role, enhancing the ability to discern fine-grained distinctions among ranked items, which is pivotal for tasks requiring precise relevance differentiation.


\begin{acks}
This work has been partially supported by Gustavus Adolphus College Faculty Startup Grants and The Fin AI community. We would like to thank the MCS department from Gustavus Adolphus College for their continuous support.
\end{acks}

\bibliographystyle{ACM-Reference-Format}
\bibliography{sample-sigconf}

\appendix



\newpage
\section{Evaluation Metrics}
\label{metrics}
\textbf{ERR} computes the likelihood that a given document will satisfy a given user query. For the ranking task, this is done by assigning each query-document pair an ``editorial grade'' between 0 and 4, with 0 meaning irrelevant and 4 meaning highly relevant. These grades are then translated into probabilities of the document satisfying the search by mapping a grade $g$ to $\left(\frac{2^g - 1}{16}\right)$. For instance, the satisfaction probability of editorial grade $3$ is $7/16$, that is, when reviewing a search result for a query with an editorial grade of $3$, there is a $7/16$ chance the user will be satisfied with that document and hence terminate the search, and a $9/16$ chance they will proceed to the next item on the ranked list. Thus, expected reciprocal rank is just the expectation of the reciprocal of the position of a result at which a user stops.  Mathematically:

\vspace{-1mm}
\begin{equation}\label{equ:ERR}
\setlength{\abovedisplayskip=0pt}
\setlength{\belowdisplayskip=0pt}
    ERR = \sum_{r=1}^{k} \frac{1}{r}  \prod_{i=1}^{r-1} (1- R_{i})R_{r}
\end{equation}

Here, $k$ is the cut-off rank for the ranked list. The probability that user stops at position $r$ is given by the term $(1- R_{i})R_{r}$.

\textbf{nDCG} applies graded relevance weighting with logarithmic discounting, preserving ordinal relevance without binarization. This ensures a robust assessment of global ranking quality, especially for long lists. Mathematically:

\begin{equation}\label{equ:DCG}
{ DCG@k}\ =\ \sum_{i=1}^{k}\frac{2^{R_i}\ -\ 1}{\log_b(i+1)}
\end{equation}

Here, $i$ denotes the position of a document in the search ranked list and ${R_i}$ is the relevance label of the $i_{th}$ document in the list, cutoff $k$ means $DCG$ accumulated at a particular rank position $k$, the discounting coefficient is to use a log based discounting factor $b$ to unevenly penalize each position of the search result. $nDCG@k$ is $DCG@k$ divided by maximum achievable $DCG@k$, also called Ideal $DCG$(\textit{IDCG@k}), which is computed from the ideal ranking of the documents with respect to the query.

\begin{equation} \label{equ:nDCG}
nDCG@k\ =\ \frac{DCG@k}{IDCG@k}
\end{equation}

\section{Relative Ranking of Methods within/across Different Tasks}\label{appen:Relative Ranking}

This experiment focuses on comparing the relative ranking of methods in terms of their $ERR$ and $nDCG$ scores, separately, \textit{within} a task or  \textit{across} different tasks (DR Vs PR).
The goal here is to see which metric yields a more stable ranking of different ranking methods across various cutoffs and types of data sets. 

We first present the ranks of different ranking methods based on their \textit{ERR} and \textit{nDCG} scores across various cutoffs. For \textit{ERR}, the performance ranking remains consistent within a specific task regardless of the cutoff. In Table ~\ref{tab:err_performance2}, we include the ranks in brackets alongside each \textit{ERR@5} score. Since the ranking is consistent across other cutoffs, thus omit the other ranks. In contrast, the performance ranking based on \textit{nDCG} varies with different cutoffs, particularly in the DR task (Table~\ref{tab:nDCG_performance2}). For instance, according to \textit{nDCG} scores, the rank of RLLaMA decreases from cutoff 5 to cutoff 60, while the rank of Gemma increases from cutoff at 5 (0.7293) to 60 (0.70517). While on PR task, ranking determined by \textit{nDCG} remains consistent across various cutoffs.

Next, we computed \textit{swap rate}~\cite{sakai2006evaluating, feng2023joint} to quantify the consistency of rankings induced by $ERR$ and $nDCG$ metrics across different data sets. The essence of swap rate is to investigate the probability of the event that two experiments are contradictory given an overall performance difference~\cite{feng2023joint}.
Table ~\ref{tab:swap_rate} shows our swap rate results for \textit{ERR} and \textit{nDCG} across two tasks.
It can be observed that nDCG (0.361) achieves a lower swap rate (high consistency) compared with ERR (0.41).

\begin{table}[!htb]
\centering
\caption{Swap rates between method rank across Document Ranking (DR) and Passage Ranking (PR) tasks. Method ranking is computed by average performance across all cutoff k}
\label{tab:swap_rate}
\begin{tabular}{l|ll}
\hline
     & \multicolumn{2}{l}{Swap Rate}  \\ \hline
ERR  & \multicolumn{2}{c}{0.41}       \\ \hline
nDCG & \multicolumn{2}{c}{0.361} \\ \hline
\end{tabular}%

\end{table}

\newpage
\onecolumn
\section{Prompt template for GPT judgment}
\label{template}

\begin{tcolorbox}[colback=lightgray!10, colframe=black, title=Prompt Template]
\fontsize{8pt}{9.8pt}
\begin{lstlisting}[breaklines=true, basicstyle=\ttfamily, frame=none]
##Instruction: You are an excellent Artificial Intelligence tasked with determining the relevance label between a given query and a passage.
##The relevance label should be selected based on the following criteria:
    5: 100% Relevant;
    4: 75% Relevant;
    3: 50% Relevant;
    2: 25% Relevant;
    1: 0% Relevant. 
##The input data consists of three components:
    query: The query to evaluate.
    passage: The candidate passage.
    binary relevance score:
        If 1.0, this means the query and passage are completely relevant. In this case, the relevance label should always be 5.
        If 0.0, you need to analyze the semantic relationship and contextual information between the query and the passage to determine the most appropriate relevance label (5, 4, 3, 2, or 1).
##Requirement:
    When predicting relevance labels, in addition to considering the semantic relevance between the query and the passage, you should also balance the distribution of your predicted labels. This means ensuring that each relevance label (5, 4, 3, 2, 1) is predicted as evenly as possible.

##Example 1:
    Input:
        query: "How to use Python for data analysis?"
        passage: "Python is a commonly used programming language for data analysis, data processing, and machine learning."
        binary relevance score: 1.0
    Output:
        {"Relevance Label": 5}
##Example 2:
    Input:
        query: "How to optimize website performance?"
        passage: "Website performance can be optimized by reducing the number of HTTP requests, and optimizing CSS and JavaScript code."
        binary relevance score: 0.0
    Output:
        {"Relevance Label": 5}
##Example 3:
    Input:
        query: "What are the benefits of regular exercise?"  
        passage: "Regular exercise can improve cardiovascular health, enhance mood, and boost energy levels. However, specific routines vary based on individual goals."  
        binary relevance score: 0.0
    Output:
        {"Relevance Label": 4}
        
##Input Format:
    query: <Insert query here>
    passage: <Insert passage here>
    binary relevance score: <1.0 or 0.0>
        
\end{lstlisting}
\end{tcolorbox}
\end{document}